\documentclass[USenglish,oneside,twocolumn]{article}

\usepackage[utf8]{inputenc}%(only for the pdftex engine)
\usepackage[big]{dgruyter_NEW}

\usepackage{graphicx}
\usepackage{url}
\usepackage{booktabs} 
\usepackage{subfigure}
\usepackage{amssymb}
\usepackage{threeparttable}
\usepackage{xcolor}
\usepackage{multirow}
\usepackage[compact]{titlesec}

\titleformat{\section}[block]{\raggedright\sffamily\selectfont\bfseries\mathversion{bold}\large}{\thesection}{1em}{}

\usepackage{algorithm}
\usepackage[noend]{algorithmic}

\makeatletter
\def\plist@algorithm{Alg.\space}
\makeatother

\begin{document}

  \author[1]{Jianwei Huang}

  \author[2]{Vinod Yegneswaran}

  \author[3]{Phillip Porras}

  \author[4]{Guofei Gu}

  \affil[1]{Texas A\&M University, E-mail: jwhuang@tamu.edu}

  \affil[2]{SRI International, E-mail: vinod@csl.sri.com}

  \affil[3]{SRI International, E-mail: porras@csl.sri.com}

  \affil[4]{Texas A\&M University, E-mail: guofei@cse.tamu.edu}

    \title{On the Privacy and Integrity Risks of Contact-Tracing Applications}
    \runningtitle{On the Privacy and Integrity Risks of Contact-Tracing Applications}

\begin{abstract}
%\boldmath
{Smartphone-based contact-tracing applications are at the epicenter of the global fight against the Covid-19 pandemic.  While governments and
healthcare agencies are eager to mandate the deployment of such applications en-masse, they face
increasing scrutiny from popular press, security companies, and human
rights watch agencies that fear the exploitation of these technologies
as surveillance tools. Finding the optimal balance
between community safety and privacy has been a challenge, and 
strategies to address these concerns have varied
among countries. 
This paper describes two important attacks that affect a
broad swath of contact-tracing applications.  The first, referred
to as {\em contact-isolation attack}, is a user-privacy attack that can
be used to identify potentially infected patients in your
neighborhood. The second is a {\em contact-pollution attack} that
affects the integrity of contact tracing applications by causing them to
produce a high volume of false-positive alerts. We developed prototype
implementations and evaluated both attacks in the context of the DP-3T
application framework, but these vulnerabilities 
affect a much broader class of applications. We found that both
attacks are feasible and realizable with a minimal attacker work factor.  
We further conducted an impact assessment of these attacks by using
a simulation study and measurements from the SafeGraph database.  Our
results indicate that 
attacks launched from a modest number (on the order of 10,000) of monitoring points 
can effectively decloak between 5-40\% of infected users in a 
major metropolis, such as Houston.}
\end{abstract}



  %\journalname{Proceedings on Privacy Enhancing Technologies}
  \journalname{}

%  \DOI{Editor to enter DOI}
  \startpage{1}
%  \received{..}
%  \revised{..}
%  \accepted{..}

%  \journalyear{..}
%  \journalvolume{..}
%  \journalissue{..}

\maketitle

% \fi
%%\IEEEpeerreviewmaketitle

\section{Introduction}

Contact-tracing applications (CTAs) have emerged as a critical tool used by many countries to combat the spread of the Covid-19 virus within their borders.   To date, more than 70 countries have launched 
contact-tracing smartphone apps. These apps rely on either Bluetooth Low Energy (BLE) or GPS to track user movements and contacts.  Many countries  have mandated the installation of these  apps on the smartphones of all of their citizens.  Among the most widely deployed tracing apps is India's Aarogya Setu application~\cite{arogyasetu} with over 50 million users in the first two weeks.  Another widely deployed application is China's contact-tracing software that uses a color-coding system (red, yellow, green) to classify citizens based on their infection risk profiles~\cite{china-app}. Russia's CTA requires its citizens to download a QR code and to declare intended travel routes to navigate through major cities, such as Moscow~\cite{russia}.   

The value proposition of CTAs is that they
provide a service to inform users when they have experienced 
close-proximity contact with another CTA user who has tested {\em
positive} for Covid-19.  When deployed ubiquitously throughout a
community, the application enables users to continually assess
their Covid-19 exposure risk.  The majority
of CTAs advertise an ability to preserve two important properties. ($i$) {\em User privacy}: the CTA
will not reveal the identity of any  Covid-19 infected CTA user. ($ii$) {\em Exposure report integrity}: the CTA service provides its users with accurate and timely exposure
reports that enable them to continually monitor their personal infection risk.

However,  the emergence of these applications has spawned significant community discussion and serious concerns regarding the extent to which these applications meet these privacy and reliability assertions. In particular, applications adopting a centralized tracking approach have
faced significant criticisms from human rights organizations and the
security community~\cite{hrw}.  An appalling example is the case of the
BeAware application from Bahrain, where users became unwitting
contestants in a television game show, with the host randomly calling
citizens and rewarding them for following social-distancing guidelines.
Other countries adopting centralized data-collection approaches include Qatar and Norway. Wen et al.~\cite{Wen20} analyzed the privacy usage in 41 CTAs and found that some applications expose identifiable information that can enable tracking of specific users. Sun et al.~\cite{Sun20} evaluated existing applications and proposed a venue-access-based contact-tracing solution which preserves user privacy while enabling proximity contact tracing.

To alleviate the privacy concerns introduced by centralized 
data-collection strategies, both Google and Apple have released alternative
CTA frameworks that have been adopted by many countries.  These frameworks espouse a fully decentralized approach, with no exchange or upload of personal information or location data, except for random numerical strings that change every 10-20 minutes.  Despite these efforts to reduce user privacy concerns, serious risks in using CTAs persist, and multiple recent studies~\cite{guardsquare, Sun20} have concluded that the vast majority lack security controls and often leak sensitive user information~\cite{Wen20}.

This paper provides a deeper examination of the privacy and integrity
properties of the dominant CTA frameworks. We describe and formalize two adversarial strategies to violate the
privacy and integrity assumptions of a wide range of CTAs, including those following the Google and Apple frameworks. The first attack, which leads to the de-anonymization of Covid-positive CTA users, is 
the {\em contact-isolation attack}.  This attack uses a combination of device spoofing, pool testing, and device re-identification techniques to efficiently narrow potential contacts to a single victim.  Pool testing has been used in various scenarios, including group blood testing~\cite{pooltesting}, de-anonymizing Internet sensors~\cite{franklin}, and most recently in waste-water-based Covid community tracking~\cite{wastewater}. We extend this concept to the problem of Covid victim identification and demonstrate that it can be implemented across communities with a modest adversary work factor. In addition, we further evaluated how attackers may leverage the Bluetooth MAC addresses obtained in contact-isolation attack to deanonymize infected users by extracting persistent identifiers (WiFi MAC, SSID) and using online SSID location databases, such as Wigle~\cite{wigle}.
  
  The second attack, {\em contact-pollution}, affects CTA integrity by polluting the Covid contact database with inaccurate contact information leading to false positive notification alerts.  The contact-pollution attack combines contact-tracing relay attacks with replay attacks.  While relay and replay attacks have been independently published in prior work~\cite{Sun20}, the specific application of this attack mix in combination with Bluetooth range extenders has not been explored.   
  
  We implemented contact-isolation and contact-pollution attacks on Android phones and evaluated them against the DP-3T (Decentralized Privacy-Preserving Proximity Tracing) framework~\cite{wiki_dp3t}.
We also assessed the implications of these attacks by using the
point-of-interest (POI) data from SafeGraph~\cite{safegraph}, which provides
aggregated and anonymized datasets on social distancing and foot traffic 
to businesses.  Such data is invaluable in tracking the spread of pandemics such as Covid.  Based on our analysis, we found that a relatively
small number of sensors (\~10,000), placed in strategic locations across a city, could be used to effectively identify devices associated with infected users.  Large-scale de-anonymization of infected individuals would also be possible if these sensors could be integrated with video cameras  and combined with face-search databases~\cite{pimeyes}.  

%{\color{red}(Maybe using a different tone could be better. For example, We formalize ..., We develop a prototype ...)}
In summary, the contributions in this paper include the following:

\begin{itemize}
    \item Description and formalization of two important attacks affecting CTAs.
    \item A prototype system that demonstrates the feasibility of both attacks and supports their evaluation against the DP-3T application.
    \item An analytical evaluation of an adversarial use of these attacks at scale,  using simulation and the SafeGraph database.
    \item Discussion of the implications, ethical considerations, and potential countermeasures to the work reported here.
\end{itemize}

\section{Background}

\label{sec:background}

To mitigate the spread of Covid-19, CTAs are being widely deployed on smartphones all over the world. These applications work by automatically exchanging data with nearby devices. When an owner of a 
cellphone with CTA is identified as infected, the CTA uploads a report to the server. The reporting process varies across applications and countries, but in many cases, it is either voluntary or a combination of self-reporting, health-authority-based reporting, and analyses of data gathered from other sources like public transportation and banking.  Infection reporting triggers contact-notification messages that will be distributed to all CTA users who may have had close contact with the patient over the past several days.  By using CTAs, users and authorities have been able to effectively track and reduce the spread of Covid-19 in many countries.

\begin{table*}
\centering
\small
\caption{\label{tab:flaws} Summary of a few published security flaws in major CTAs.}
\begin{tabular}{cccp{10 cm}} %\hline
App Name & Country & Architecture & Published Security Flaws \\ \hline

EHTERAZ & Qatar & Centralized & A QR-code vulnerability was discovered that leaked information potentially allowing hackers to harvest more than a million people’s national ID numbers and health status. The app is reported to be mandatory in Qatar with three years in prison and a fine of QR 200,000 (about \$55,000) for non-compliance~\cite{etheraz-flaws}. \\ \hline

BeAware & Bahrain & Centralized & \multirow{3}{10cm}{These contact-tracing apps export real-time GPS tracking information to a central database which might be used for surveillance~\cite{amnesty-flaws}.  Norway has stopped the development of its CTA due to pushback.}\\
Shlonik & Kuwait  & Centralized & \\
Smittestop & Norway & Centralized & \\ \hline

%https://www.wired.com/story/india-Covid-19-contract-tracing-app-patient-location-privacy/
Aarogya Setu & India & Centralized & Researchers discovered a vulnerability in the
app feature to discover infected people who were nearby (within a 500 m
radius) using GPS coordinates. The authors demonstrated a
triangulation attack that allows tracking infections to individual
homes in sparse locations and confirm Covid status of suspected
individuals. The app is mandatory for moving around in India~\cite{aarogya-flaws}. \\ \hline

%https://www.bbc.com/news/technology-52725810
NHS App & UK & Decentralized &  Several vulnerabilities were identified including lack
of TLS certificate validation, use of long-lived BLE BroadcastValues, device reidentification
attacks based on the use of a deterministic keep-alive counter, storing unecrypted data
on handsets that could be used by law-enforcement to determine contacts etc~\cite{nhs-flaws}. Based on these reports, UK decided to switch to a backup option that uses a decentralized model.\\ \hline

Covid Safe & Australia & Decentralized & Both of these applications use the OpenTrace framework which uses the\\
TraceTogether & Singapore & Decentralized &  BlueTrace protocol.  These are vulnerable to long-term device-re-identification attacks due to the fact that tempIDs are used for extended periods and may present themselves across locations~\cite{opentrace-flaws1, opentrace-flaws2}. These vulnerabilities could be exploited in tandem with our contact-isolation attack.\\ %\hline

\end{tabular}
%\vspace{0.1in}
\end{table*}

\subsection{Anatomy of Contact-Tracing Applications}
\label{sec:anatomy}
\begin{figure*}[htbp]
    \centering
    \includegraphics[width=0.9\textwidth]{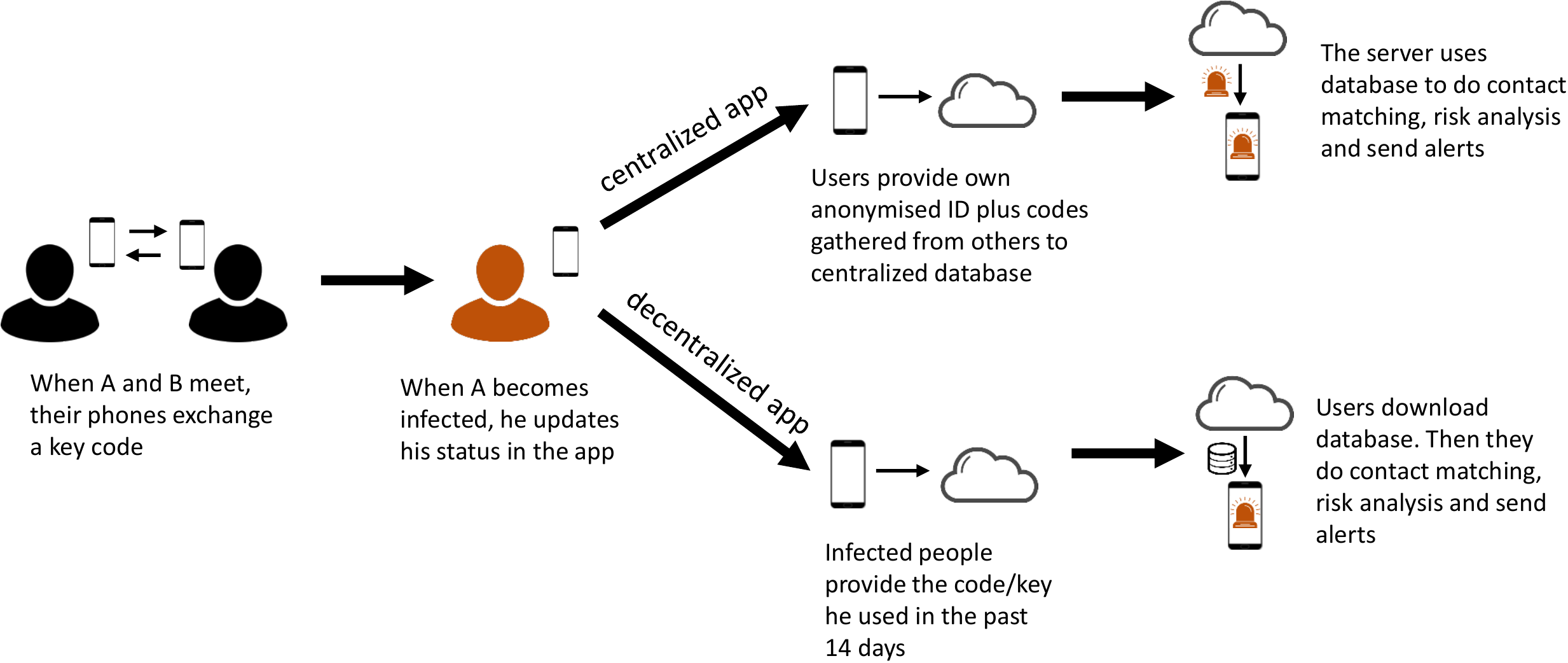}
    \caption{Illustration of the messaging workflow in centralized and decentralized architectures used by Covid-19 CTAs.}
    \label{app_arch}
\end{figure*}

The high-level architecture of a typical CTA is illustrated in Figure \ref{app_arch}. When two users, A and B, become proximate, their phones will exchange a key code. The code is generated from the private information of users using advanced cryptographic algorithms (e.g., RSA, SHA-1). Most applications use the advertisement message in Bluetooth Low Energy (BLE) to exchange data, which persistently broadcasts messages to nearby devices. If user A subsequently becomes infected, he updates his status in the application. The design of the application determines which of the two possible strategies will be used for distributing the notification messages.

\vspace{0.05in}
\noindent \textbf{Centralized Applications.} For centralized applications, user A provides his anonymized ID plus codes gathered from other proximate phones to the centralized server. The server decrypts these codes or does contact matching in the database to identify all potentially affected users and then alerts them.

\vspace{0.05in}
\noindent \textbf{Decentralized Applications.} For decentralized applications, user A provides the code/key used in the past 14 days to the server. All users of the application periodically download the database and perform local contact matching. In this case, alerts are sent from client-side applications installed on user devices.

Table~\ref{tab:flaws} provides a summary of a few of the major published security flaws in CTAs.  In most cases, governments have been quick to respond to published flaws, by either issuing security patches, rethinking architectural designs, or scrapping the application.  While security flaws seem more egregious and pervasive in centralized architectures, they affect both classes of applications.  None of these disclosures specifically discuss the attacks that are described in this paper, however, these currently documented adversarial techniques could be combined with the attacks presented here to increase their effectiveness.  For example, the GPS-based triangulation attack could be combined with the Bluetooth-based contact-isolation attack.

\subsection{Data Exchange Protocol}

As mentioned in ~\ref{sec:anatomy}, most CTAs will communicate with nearby devices to record close contacts. To satisfy the privacy and feasibility requirements of CTAs, the data-exchange protocol should be easily implementable and able to guarantee the anonymity of exchanged messages. Hence, Bluetooth, as one of the most widely used wireless communication protocol, is used in almost all the contact tracing applications as the data-exchanging protocol. Specifically, CTAs leverage the advertisement message to actively send generated code to nearby devices. Meanwhile, the applications will continue collecting advertisement messages and extract the code of nearby devices from these messages. The benefit of advertisement messages is that no device-specific information is included in the message.

\begin{figure}[htbp]
    \centering
    \includegraphics[width=\columnwidth]{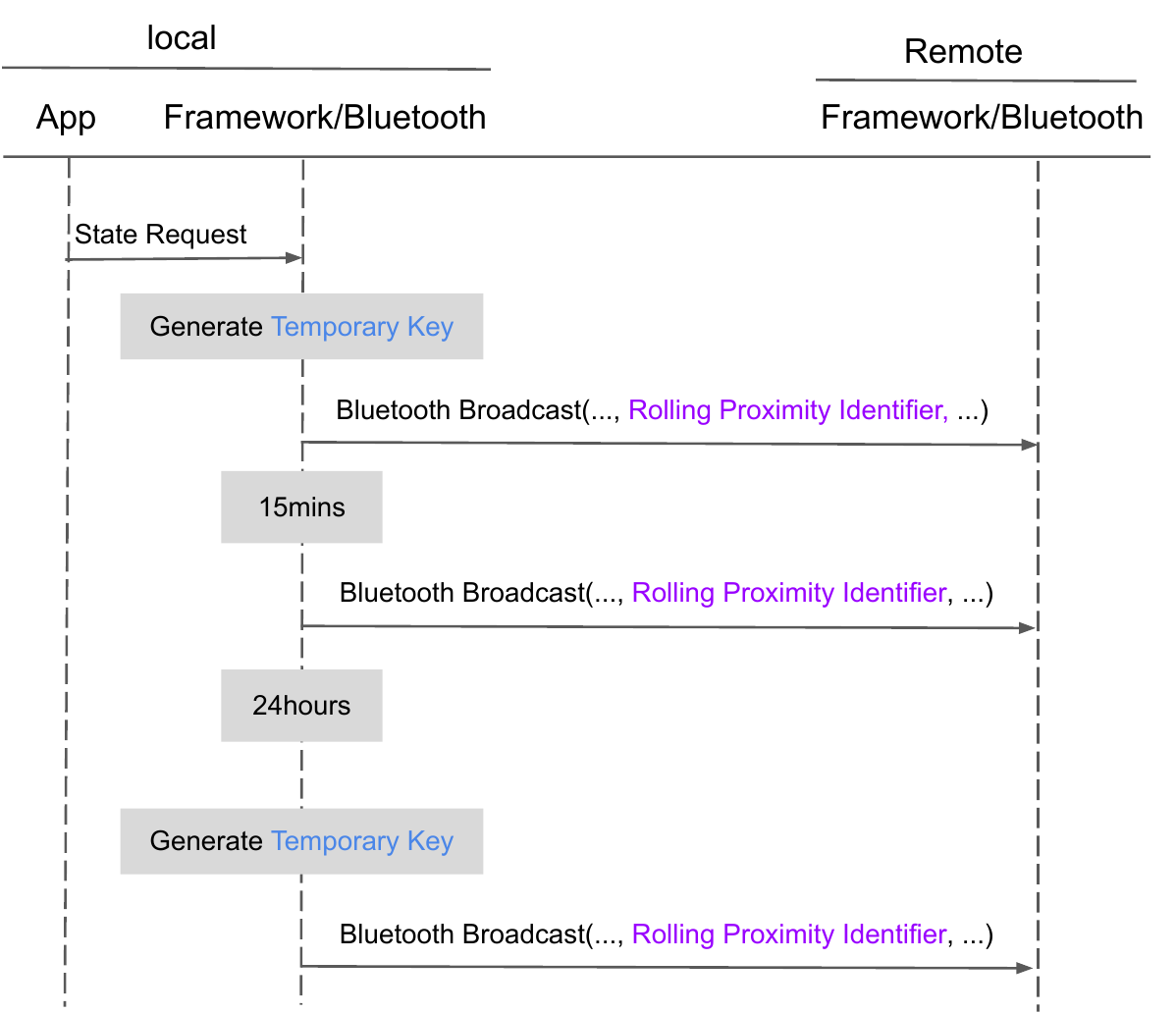}
    \caption{Data Exchange Protocol used by Google and Apple Frameworks.}
    \label{comm_proto}
\end{figure}

\begin{table}[!htbp]
    \small
    \centering
    \caption{Inputs for exchanged code-generation functions.}
    \label{tab:encounter}
    \resizebox{1\columnwidth}{!}{
        \begin{tabular}{ccccc}
            %\toprule
            Application  & Temporary Key & Timestamp & Location & Algorithm\\
            \midrule
            tracetogether & \checkmark & \checkmark & \checkmark & AES\\
            mytrace   & \checkmark & \checkmark & & AES\\
            aarogya   & \checkmark & \checkmark & & AES, SHA-256\\
            rakning   &  & \checkmark & & AES, SHA-256\\
            erouska   & \checkmark & & & HKDF, AES\\
            Covidsafe  & \checkmark & & \checkmark & AES\\
            geohealthapp & & \checkmark & \checkmark & AES\\
            novid  & \checkmark & & & AES\\
            stop corona & \checkmark & & & HKDF, AES\\
            smittestop & \checkmark & & & AES\\
            hamgen & \checkmark & & & AES\\
            swissCovid(DP-3T) & \checkmark & & & HKDF, AES\\
            %\bottomrule
        \end{tabular}
    }
\end{table}

As an example, the detailed workflow of the sender part of the data-exchange protocol in Google and Apple contact tracing frameworks is shown in Figure \ref{comm_proto}. When the local CTA requests to send the state code to nearby devices, the framework first generates a temporary key. Each time the framework sends an advertisement message, a rolling proximity identifier will be generated from the temporary key. The default interval of updating the identifier is 15 minutes. 
%According to Bluetooth specifications, the maximum length of advertising payload is 31 bytes. In practice, the length of rolling proximity identifier is 16 bytes.
Since the identifier generation algorithm is open-sourced, it is possible to associate different identifiers and further identify the device with enough identifiers. To mitigate the security issue, the temporary key is updated every 24 hours.

Different applications generate the exchanged code with different inputs. We collected 12 popular CTAs in different countries and analyzed the information used (see Table \ref{tab:encounter}). Most applications  generate a temporary key periodically and use an advanced cryptography algorithm to generate the code.

\section{Contact-Tracing Attacks}

Central to the appeal of CTAs is the premise
that they provide users a timely and accurate risk assessment service
while also protecting each user's personal Covid-19 infection status.
More specifically, CTAs typically advertise an ability to preserve two
important properties: ($i$) {\em User privacy}, which ensure that each
user's Covid-19 status is protected from other users, and ($ii$) {\em
Exposure risk integrity}, which ensures that users receive timely and
accurate risk indicators when near-proximity exposure to other
Covid-19-infected CTA users occurs.  One may observe that these two
properties are mutually contradictory, as the satisfaction of one
property  will likely prevent satisfaction of the other
property.  For example, if a sequestered CTA user meets only one
other person during a given week and then subsequently receives a
Covid-19 exposure report, then it is trivial to infer that the other user
is Covid-19 positive.  However, withholding the exposure report to avoid
such a de-anonymization would violate the requirement for accurate and
timely report delivery to ensure that the sequestered user can
properly assess their personal risk.

Even when considered separately, we find that adversarial
models exist that directly violate each of these two properties across
multiple CTA frameworks.  In this section, we discuss these models and 
present specific methods to implement attacks within a reasonable 
adversary work factor.  
The first attack, (i.e., {\em contact-isolation attack}) enables
one to selectively violate the user privacy property of CTAs.  
The second attack (i.e., {\em
contact-pollution attack}) enables adversaries to prevent CTAs
from satisfying the exposure risk integrity property. These attacks are 
quite general, and affect a broad class of
applications, and are agnostic to whether the CTA follows the
centralized or decentralized model.

% In this section, we first describe our threat model. Then we provide  details of the two attacks: the contact-isolation attack and the contact-pollution attack.  The specific attacks that we describe are quite general and affect a broad class of applications following both the centralized and decentralized model. 

\subsection{Threat Model}
Our threat model assumes that attackers have full access to their own devices. Attackers can modify the application stack running on phones, Bluetooth drivers of their devices, data exchanged with other devices, and data uploaded to remote servers from their devices. However, attackers have no access to other user devices and servers.  We also make no assumptions about the ability of attackers to exploit remote servers or other user devices.

The model further assumes that attackers can fully reverse-engineer the workflow of the clients of target applications by analyzing the source code or binary file of the clients of target applications. Attackers know how to ($i$) create multiple accounts; ($ii$) upload records to servers; ($iii$) exchange data with other devices; and ($iv$) check notification messages. We do not  consider the applications if  strict restrictions are deployed on the server side to stop attackers from emulating these four operations. Although attackers know the workflow of clients, we assume that all communication between different user devices and all communication between devices and servers are well-protected by HTTPS and Bluetooth protocols.%[FIX ME] footnote about how may bypass some restrictions (e.g. ip blocking rule).

For automated large-scale de-anonymization, we assume that these sensors are placed at popular locations (cafes, gas stations, etc.) in the city.  We further assume that these sensors have integrated video
cameras for effective device-to-victim persona mapping.

\subsection{Attack Formalization}
To clearly illustrate the attacks, we first formalize the workflow of typical CTAs,  assuming an attacker is the user of the target application. When the attacker has close contact with other users, each of these devices will mutually exchange anonymized data. The data distributed by the attacker's device is indicated as $c_{att}$, and the data that the attacker's device receives is $C_{att}=\{c_1, c_2, c_3, ..., c_n\}$. The test result of the attacker is $r_{att}$. By default, $r_{att}$ is $false$. Similarly, the set of received test  results is $R_{att}=\{r_1, r_2, r_3, ..., r_n\}$ and $r_i=false$, where $i=1, 2, 3, ..., n$.

A user $k$ who is confirmed to be infected,  will report their infection to the server, which sets $r_{k}$ to $true$. If the attacker had recent close contact with user $k$, $d_k \in D_{att}$ and $r_k=true$. Now, the attacker is considered to be at risk for the infection. In this case, notification messages, denoted as $N$, will be sent to the attacker. Here, $N$ is $\Sigma {R_{att}}$.

\subsection{Contact-Isolation Attack}
By abusing notification messages, the contact-isolation attack allows an attacker to identify the device whose owner is confirmed Covid positive, which breaks the anonymity of CTAs. 

\begin{figure*} 
  \centering 
     \includegraphics[width=1.0\textwidth]{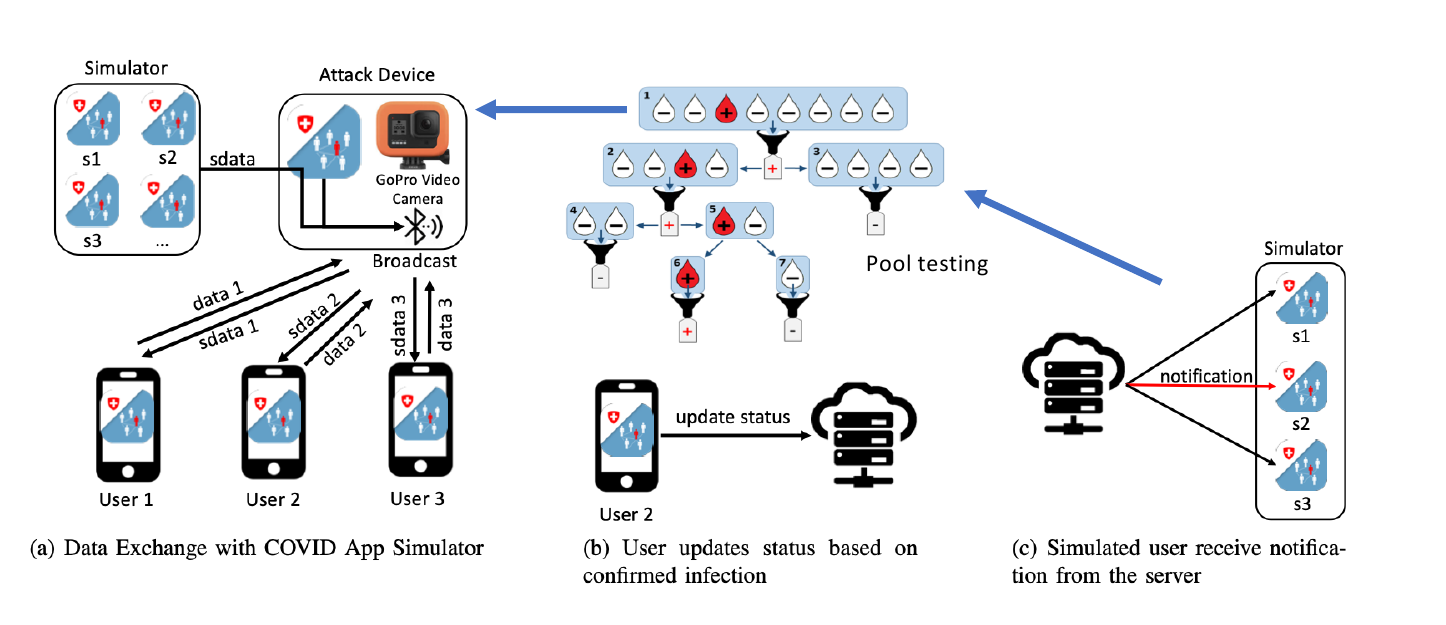}
  \vspace{-0.2in}
  \caption{Illustration of the contact-isolation attack. Many applications are simulated in the device simulator. The attack device exchanges the data from simulated devices with normal devices, monitors 802.11 activity, as well as captures a video feed of visitors entering the location. When normal user updates his status in the server, only the simulated application whose data is exchanged with the confirmed case will receive the notification message.} 
  \label{fig:attsimulation} %% label for entire figure 
\end{figure*}

When the devices of confirmed cases in $R_{att}$ are $R_{att\_confirmed}=\{r_{c1}, r_{c2}, r_{c3}, ..., r_{cm}\}$, the contact-isolation attack aims to divide $C_{att}$ into many sublists $C_1, C_2, C_3, ..., C_s$. Ideally, for each confirmed case $r_p \in R_{att\_confirmed}$, if $r_p \in C_q$, $r_p \notin C_k$, where $1\le p\le m$, $1\le q\le s$, $k=1,2,3,...,s$ and $k \ne q$.

Then the attacker creates $s$ new accounts. For each new account $u_i$, the attacker uploads $C_i$ to the server. If the notification message is sent to $u_i$, there must be at least one confirmed case in $C_i$. If $|C_i|$ is 1, the device is exactly the one that belongs to the confirmed case. If not, by repeating the approach, the attacker can finally reduce $|C_i|$ to 1 and find the device.

\subsection{Contact-Pollution Attack}

False positive (and negative) alerts are a significant
concern for most security applications and healthcare
assessment tests. The same holds true for CTAs. For example, 
%{\color{red}(Maybe you should not refer to a paper this way. You should say "For example, Schneier et al. [Reference] observers that ...)} in
Schneier et al.~\cite{Schneier} observe that CTAs typically 
define a risk exposure based on a six-foot proximity
of two CTA users for more than ten minutes. Unfortunately, the dependencies on GPS and Bluetooth signaling are imprecise due to a range of factors
that are inherent with RF propagation or mobile GPS
signal inaccuracies or location update delays.

While our paper refrains from making any such high-level determinations on the utility of Covid CTAs, the key takeaway is that high false positive or false negative rates can render these applications useless.  This section describes how a contact-pollution attack corrupts the contact database of target applications and results in many false positive notification messages. Increasing the false positive rate of a contact tracing application will decrease its reliability initially making them  paranoid and subsequently cavalier to alerts produced by the system.

Existing CTAs assume that all the data from other devices is trustworthy because they use advanced cryptographic algorithms to generate the exchanged data. However, since we can both send and receive data, there are two avenues for the attacker to insert fake contact data that pollutes the contact database.

\begin{figure}[tb]
    \centering
    \includegraphics[width=0.9\columnwidth]{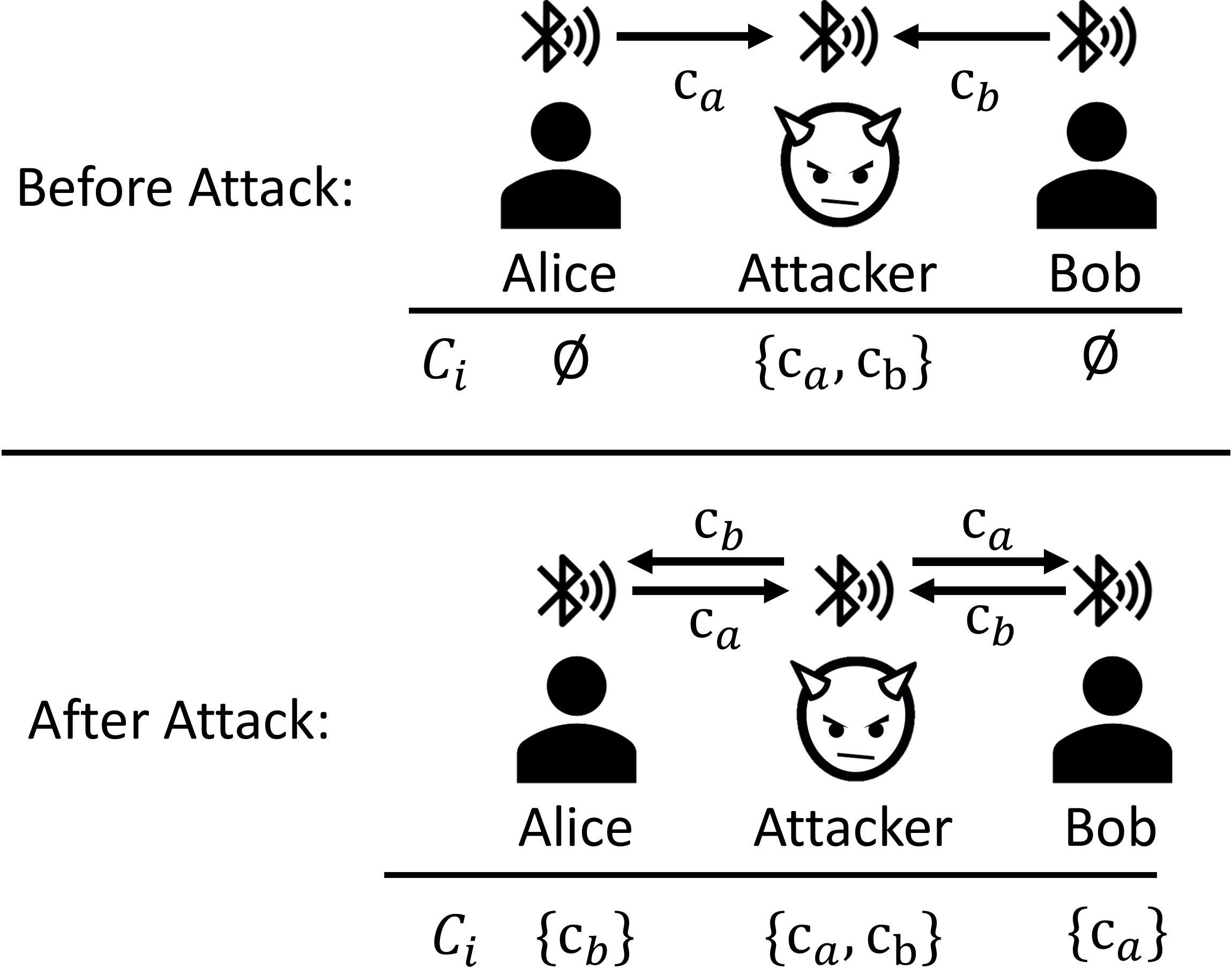}
    \caption{Example of contact-pollution attack. Both Alice and Bob have close contact to attacker but they are out-of-range of each other. However, when the attacker impersonates them, Alice and Bob will receive the data from each other and think they had close contact with each other.}
    \label{contact_polluting_exp}
\end{figure}

\begin{enumerate}
    \item The attacker can send the data received from other devices to all the devices in range; i.e., attacker sends $C_{att}$ to $d_i$ where $d_i$ is in range of attacker's device. As shown in Figure \ref{contact_polluting_exp}, when Alice and Bob are in range of the attacker's device, but not in mutual range, they will still receive fake contact data from each other.
    
    \item If the attacker has multiple devices in different places, these devices can communicate with each other and share the received data. Thus, $C_{att}$ contains the data we received through all attacker's devices, and the attacker keeps sending $C_{att}$ to all other devices in range of his devices. In this case, when Alice is shopping while Bob is attending classes, if they are in range of attacker's devices, they will receive fake contact data from the attacker, who is impersonating both of them.
\end{enumerate}

While the first approach effectively doubles the attack range, the second approach infinitely extends it. Both of these approaches will result in additional $r_k$ in $R_{Alice}$ and $R_{Bob}$. Thus this attack can stimulate a large number of fake notification messages when $r_j=true$ is inserted into $R$ of other users. 
\section{Launching Attacks in the Real World}
\label{sec:implementation}

This section describes how the previously described attacks can be implemented. As discussed in Section~\ref{sec:background}, most CTAs leverage Bluetooth to collect contact information, therefore, 
our attacks are primarily focused on the Bluetooth medium.
We use the DP-3T framework as an example to illustrate how we can apply 
the attacks to real-world CTA scenarios.

\subsection{Contact-Isolation Attack}
\begin{figure}[htbp]
    \centering
    \includegraphics[width=1.0\columnwidth]{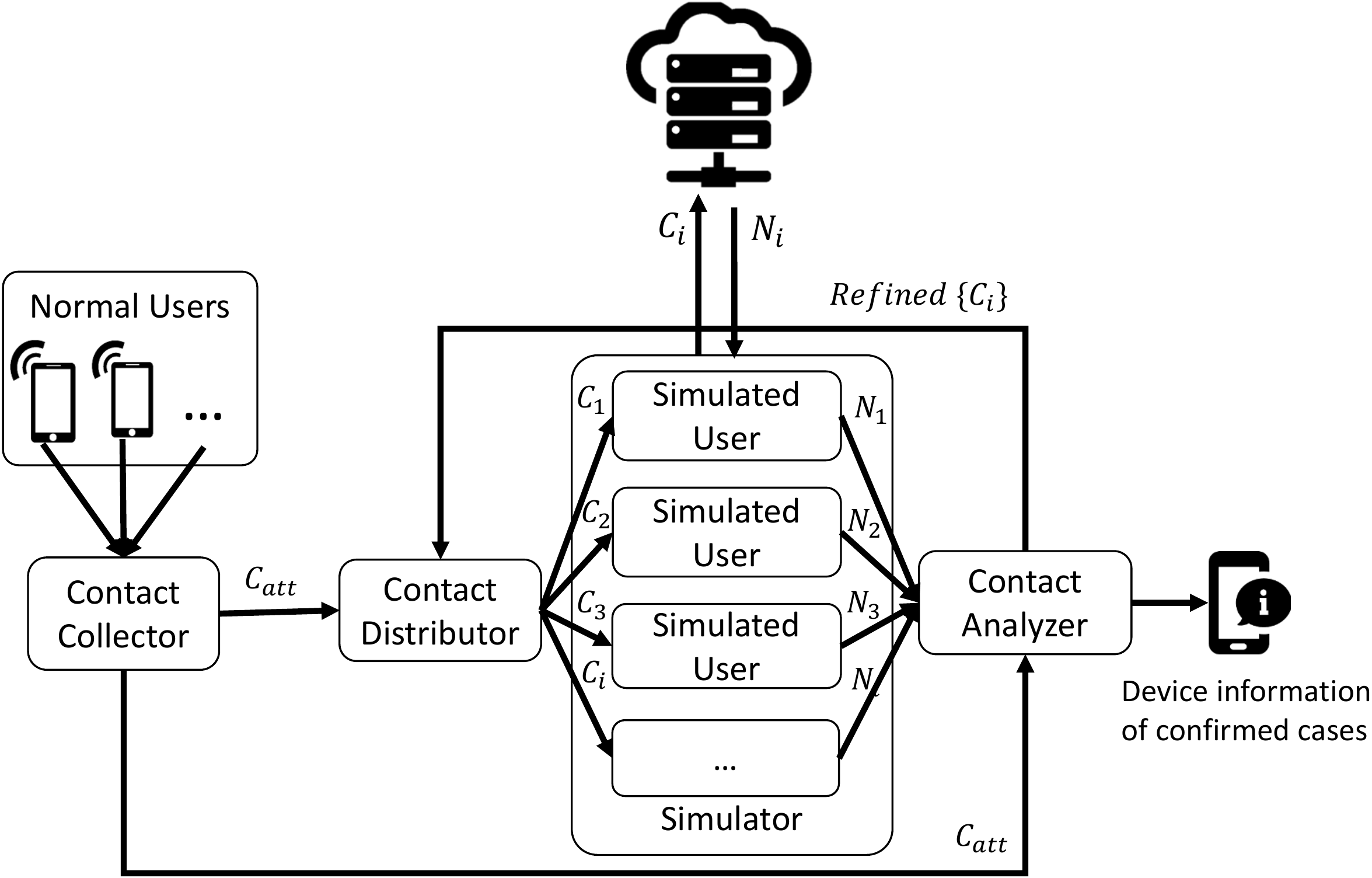}
    \caption{Prototype implementation of the contact-isolation attack.}
    \label{isolation_arch}
\end{figure}

The architecture of our prototype implementation for the contact-isolation attack is illustrated in Figure. \ref{isolation_arch}. Our implementation is divided into four components. The first component,  the \emph{Contact Collector}, is deployed on the attackers' devices
and is used to exchange data with normal CTA users.  The second component, the \emph{Contact Distributor}, separates the collected data into several groups and sends them to the \emph{Device Simulator} based on the algorithm illustrated in Algorithm \ref{alg:contactanalyzer}. The initial value of $\{C_i\}$ is $C_{att}$, so we divide $C_{att}$ into $n$ parts to produce ${C_i}$ for the first run. Each Device Simulator will have a unique account created by attackers. When the Device Simulator receives the contact data, it will upload the data to the server and wait for notification messages. All received notification messages will be forwarded to the Contact Analyzer, which extracts the device information of confirmed cases.

\vspace{0.05in}
\noindent \textbf{Contact Collector.} To collect the contact information, the target application will be running as a normal user. We use Frida~\cite{frida} to dynamically instrument the target application. Since most CTAs are based on Bluetooth, we hooked the Bluetooth API to intercept the beacon data from other devices before they are handled by the application. More specifically, we hooked \emph{MessageListener.onFound} on Android.

\vspace{0.05in}
\noindent \textbf{Contact Distributor.} When $C_i$ is sent to the Contact Distributor, $C_i$ is divided into several subsets and sent to different simulated devices to communicate with the server.

\vspace{0.05in}
\noindent \textbf{Device Simulator.} The Device Simulator creates a set of simulated devices. Each device will maintain a simulated user and the communication between the device and the server. Our implementation extracts the relevant logic of the target application and repackages it. So each simulated device is actually a process running on our device. The challenge in the Device Simulator is how to create the necessary number of accounts, as most CTAs require personal information to sign up, however, only a few required personal attributes are verified. Most applications only verify the phone number, which can be easily obtained from an online temporary phone number service. Since the number of accounts needed for most cases is under 30, we can manually register the accounts for our attacks.

DP-3T does not have difficult restrictions for account creation. For confirmed infection cases, users are identified by a Covidcode, which is assigned to users by the government or hospital. Other users do not need to upload any information to the server. The only challenge in launching the contact-isolation attack is the possible firewall deployed on the server side, which is not included in the application itself. Therefore, we deployed DP-3T on our local machine with a simple firewall, which rejects connections from the same IP address when the number of connections reaches the threshold (10 in our evaluation).

\vspace{0.05in}

\noindent \textbf{Contact Analyzer.} When all the $N_i$ have been collected, the Contact Analyzer will check if one of them is $true$. If so, the Analyzer will refine sets of $C_i$ for simulated users to get another set of $N_i$. The algorithm to refine $C_i$ is as shown in Algorithm \ref{alg:contactanalyzer}.

\begin{algorithm}
\caption{Contact-Refinement Algorithm}
\label{alg:contactanalyzer}
\small
\begin{algorithmic}[1]
    \REQUIRE $\{C_i\}$, where $i=1,2,3,...,n$ used in last run, corresponding status $\{N_i\}$ and number of simulated devices/applications $n$.
    \ENSURE refined $\{C'_i\}$
    
    \STATE ${C'_i} = \emptyset$
    \STATE $T = \emptyset$
    \STATE $positive\_counts = 0$
    \FOR{$k$ from $1$ upto $n$}
        \IF{$N_k == true$}
            \STATE $positive\_counts = positive\_counts + 1$
            \STATE $T = T \bigcup {N_k}$
        \ENDIF
    \ENDFOR
    
    \STATE $set\_pool = n / positive\_counts$
    \STATE $counts = 0$
    \FOR{$N_p \in T$}
        \STATE $set\_size = |N_p| / set\_pool$
        \FOR{$j$ from $1$ upto $set\_pool$}
            \STATE $C'_{counts*set\_pool+j} = $\{the (set\_size*j)th to (set\_size*j+j)th elements in $N_p$\}
        \ENDFOR
        \STATE $counts = counts + 1$
    \ENDFOR
\end{algorithmic}
\end{algorithm}

\vspace{0.05in}
\noindent {\bf Scenario 1: Controlled-Encounter Scenario.} %Frequent Contact Deanonymization Attack}.
In this scenario, the target set includes frequent contacts (e.g., coworkers, neighbors).
The attacker walks around (the neighborhood, apartment, dorm, office
building) choosing a different (potentially overlapping) set of houses
or offices in each round. In each round, the attacker spoofs a
different device identifier.  By repeating this experiment for	a few
rounds and using pool testing, the attacker can quickly narrow down to
a small number of infected neighbors.  The key point is that the
attacker can have a potentially unlimited number of encounters
with each contact since	they reside in the same	community.  In addition,
Bluetooth range extenders also improve the feasibility of the attack,
as interactions	no longer have to be close physical interactions.

%{\em What is the adversary work factor?}
If the attacker creates \emph{n} accounts for pool testing, and the number of users of the application in the community is \emph{N}, the attacker only needs to do $log_n N$ rounds of pool testing to identify all the infected users.

\vspace{0.05in}

\noindent {\bf Scenario 2: Opportunistic-Encounter Scenario.} %Automated Large-Scale Harvesting.}
This scenario requires three additional capabilities for uncontrolled de-anonymization of infected users
to deal with the challenges involved in scaling and device re-identification.  
First, we address scaling by using a pool-testing strategy that involves reusing a set of identifiers across groups of contacts. 
Second, we will track WiFi broadcast activity in addition to Bluetooth activity, and use temporal correlations to map WiFi addresses with Bluetooth activity.  While recent Android and iOS versions implement MAC address randomization,
this happens only during the probe stage and per-SSID MAC consistency is maintained to facilitate automatic
re-authentication. In addition, automatic WiFi probes also broadcast the names of recent networks to which a device 
has connected to that leak additional information, such as names of home or work networks, airports, cafes visited, etc.
This information can then be used or combined with other device re-identification attacks~\cite{opentrace-flaws2} to track repeated visits by users that 
help to narrow the precision from groups of devices to individual devices.  Thus, 802.11 (SSID, MAC) pairings provide a means to go from an ephemeral identifier (Bluetooth MAC) to a persistent identifier.  
Third, information from the integrated video camera could be used to derive the mapping from device identifiers to individuals.
In addition, private SSID information can be used for location tracking of individuals using public databases, such as Wigle~\cite{wigle}.

To associate the WiFi probes to Bluetooth addresses, we differentiate the pair of probes and addresses by checking their existence and the time we first detected them. Since the signal strengths of WiFi and Bluetooth are different, we may not be able to identify all pairs if they only appear once. That is, for most devices the attacker will be able to associate the WiFi probe and the Bluetooth address if it communicates multiple distinct times (encounters) with the attack device. Based on our analysis in Section~\ref{sec:de-ano}, most pairs can be identified with two encounters.  Each visit to a monitoring location (e.g., store, cafe) would likely result in multiple encounters.

\subsection{Contact-Pollution Attack}
\begin{figure}[htbp]
    \centering
    \includegraphics[width=0.9\columnwidth]{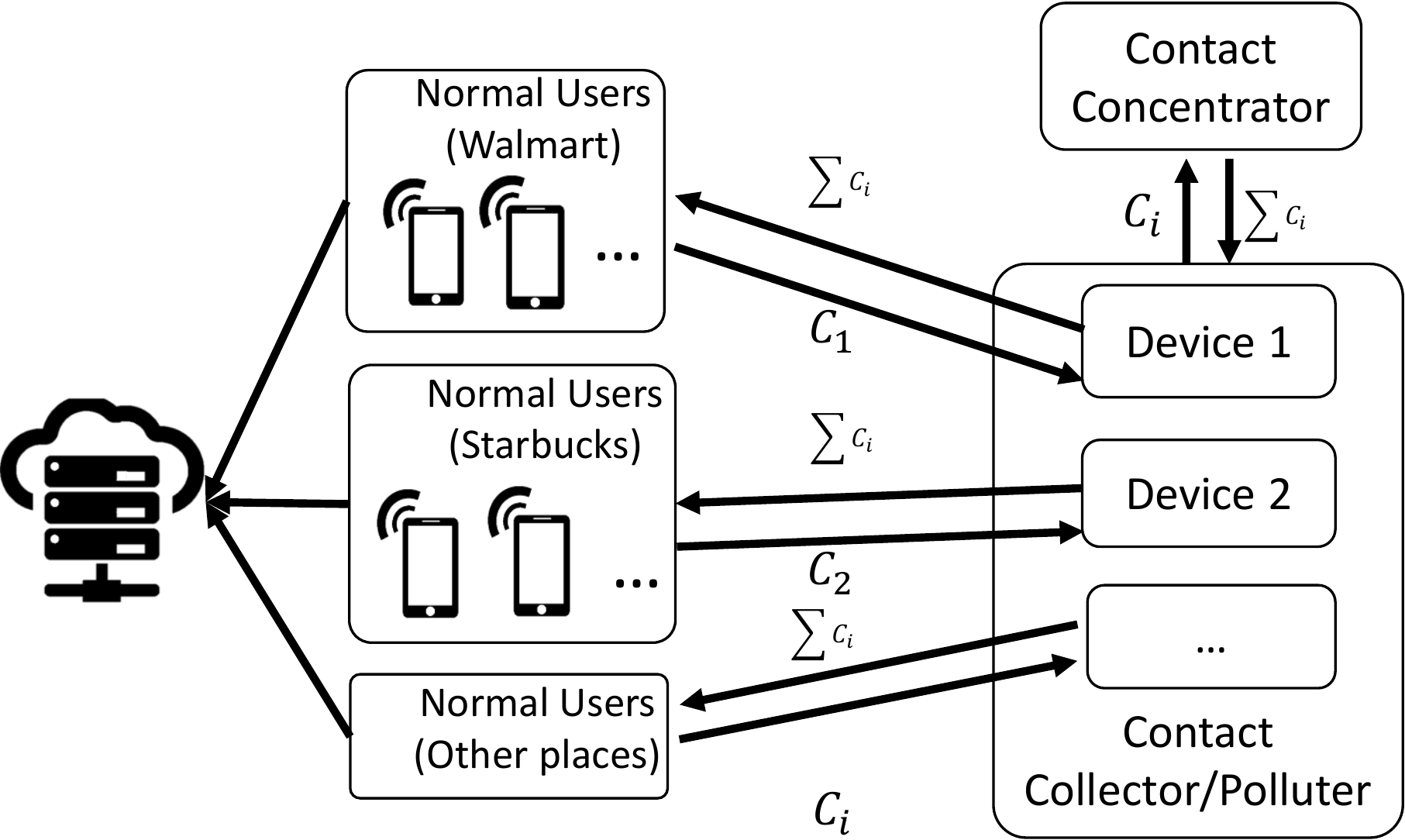}
    \caption{Prototype implementation of the contact-pollution attack}
    \label{contact_polluting_arch}
\end{figure}

The prototype implementation of contact-pollution attack has three main components, as illustrated in Figure \ref{contact_polluting_arch}. We continue to use the \emph{Contact Collector} to gather contact information from the devices of other users. All collected contact information will be transmitted to the \emph{Contact Concentrator}. Once the Contact Concentrator receives new data, all the \emph{Contact Polluters} will receive a new database.  The Contact Polluters use this information to continuously broadcast as many contact messages as they can.\footnote{\noindent The actual number of messages in practice is limited due to the physical limitation of devices. Details will be described in Section \ref{sec:attasimulation})}

\vspace{0.05in}

\noindent \textbf{Contact Concentrator.} The Contact Concentrator is implemented as an HTTP server, which provides two API functions for the Contact Collector and Contact Polluter, respectively, to upload and download contact information.

To further spread false contacts, attackers need to collect the contact information from the original user and replay it to another user. To introduce $n$ false contacts to the database, attackers should participate in at least $2n$ interactions with other users.
\section{Data-Driven Impact Assessment}
\label{sec:eval}

This section presents our evaluation results that focus on the practicality and impact of the two attacks.
More specifically, our evaluation seeks to answer the following research questions:

\begin{itemize}
    \item[R1.] How feasible are the proposed attacks?
    \begin{itemize}
        \item[I.] Are the attacks practical for real-world applications?
        %\item[II.] What is the upper bound of the number of victims for each attack device?
        \item[II.] How much time do attackers need to complete the attacks?
    \end{itemize}
    
    \item[R2.] How can the attacks affect the users of CTAs?
    \begin{itemize}
        \item[I.] Is it possible to de-anonymize the victims? 
        \item[II.] How many users are at risk?
    \end{itemize}
\end{itemize}

To answer R1.I, we evaluated the challenges in launching the attacks on real-world applications (Section \ref{sec:attasimulation}). We also analyzed the limitations and performance of each attack to determine the answer to R1.II.
For R2.I, we tested the existing techniques of identifying users with Bluetooth MAC address 
(Section~\ref{sec:de-ano}). The results show that user identities can be leaked to attackers in specific cases. Finally, we empirically evaluated the number of possible victims in different cities with the database from SafeGraph to answer R2.II (Section~\ref{sec:empirical}).

\subsection{Data Collection}
To answer the first set of questions, we evaluated our attacks on the sample application and the backend of DP-3T. We simulated the attacks with different victim devices and different attacker devices. The threshold of distance that devices will exchange data is 5 meters, and the time threshold is set to 1 second to simplify the evaluation process.

We obtained POI datasets from SafeGraph. The datasets contain addresses, open hours, visitor counts, and popular times of over 1 million places. We leveraged these datasets to explore the question of large-scale attack deployment strategies that could maximize the attack coverage across city populations. We performed further analyses regarding de-anonymization based on the results of our attacks. This paper uses Houston and Bryan, TX and Chicago, IL as examples to show how the attacks can possibly affect the users of CTAs.\footnote{For these analyses, we assume that CTAs are mandatory to everyone and have been deployed on all user devices.}

\subsection{Attack Simulation}
\label{sec:attasimulation}

To simulate the attack on the sample application of DP-3T, we deployed the server on AWS. The devices we used in the simulation are shown in Table \ref{tab:devicelist}. For iOS devices, we made some modifications to the source code of the client to avoid jailbreaking the devices. The modifications include the following elements. ($i$) When a device exchanges data with another device, the data and information of the other device will be stored into local file. ($ii$) All contact-tracing-related operations (i.e., broadcast data, retrieved data,  data uploaded to the server) will be logged with timestamps.

\begin{table}[t]
\small
\centering
\caption{List of devices used for the simulation experiment}\label{tab:devicelist}
\begin{tabular}{ccc}
%\toprule
Device  & Model & OS version\\
\midrule
A & iPhone 11 & iOS 14.0\\
B & iPhone SE (2nd ed.) & iOS 13.5.1\\
C & Xiaomi Mix 2 & MIUI 10 Global 9.6.27\\
D & Macbook Pro (2020) & MAC OS X 10.15.5\\
%\bottomrule
\end{tabular}
\end{table}

For Android devices, we use Frida to instrument the application and implement functionalities analogous to those we implemented on iOS devices. To store the data and device information of another device, we hooked the {\em MessageListener.onFound} API function and extracted the parameters sent to it. To log all contact tracing related operations, we looked into the source code of the sample application and found corresponding methods (Table \ref{tab:hookedmethods}) that were hooked to log operations.

\begin{table}[t]
\small
\centering
\caption{Required information for creating account in CTAs}
\label{tab:required_info}
\resizebox{1.02\columnwidth}{!}{
\begin{threeparttable}
\begin{tabular}{ccccc}
%\toprule
Application  & Phone Number & Valid ID\tnote{*} & Device ID\\
\midrule
tracetogether   & \checkmark & \checkmark & \checkmark\\
mytrace   & \checkmark & &\checkmark\\
aarogya   & \checkmark & &\checkmark\\
rakning   & \checkmark & & \checkmark\\
erouska   & & &\checkmark\\
Covidsafe  & \checkmark & & \checkmark \\
geohealthapp & & & \checkmark \\
novid  & & & \checkmark \\
stop corona & & & \checkmark \\
smittestop & & & \checkmark \\
hamgen & & & \checkmark \\
swissCovid(DP-3T) & & & \checkmark \\
%\bottomrule
\end{tabular}

\begin{tablenotes}
\footnotesize    
\item[*]{passport, driver license, medical ID, etc.}
\end{tablenotes}
\end{threeparttable}
}

\end{table}

\begin{table}[t]
\small
\centering
\caption{Summary of recorded encounter data in devices}
\label{tab:recordeddata}
\begin{tabular}{ccc}
%\toprule
Device Name & Encounter & Recorded Encounter\\
\midrule
$V_A$ & 8B13933B4E21... & {37B1BA608307...}\\
$V_B$ & 8B13C29438F0... & {37B1F8BE1FE9...}\\
$V_C$ & 8B13048BF17D... & {37B18307AB63...}\\
\midrule
$A$ & 37B1BA608307... & 8B13C29438F0...\\
~ & ~ & 8B13C29438F0...\\
~ & ~ & 8B13048BF17D...\\
\midrule
$B$ & 37B1F8BE1FE9... & 8B13C29438F0...\\
~ & ~ & 8B13C29438F0...\\
~ & ~ & 8B13048BF17D...\\
\midrule
$C$ & 37B18307AB63... & {8B13C29438F0...}\\
~ & ~ & 8B13C29438F0...\\
~ & ~ & 8B13048BF17D...\\

%\bottomrule
\end{tabular}
\end{table}

\begin{table*}[t]
\small
\centering
\caption{Summary of operation methods hooked in Android Applications}\label{tab:hookedmethods}
\begin{tabular}{ccp{7cm}}
%\toprule
Class  & Method & Description\\
\midrule
GoogleExposureClient & getTemporaryExposureKeyHistory & Retrieves key history from the data store on the device\\
GoogleExposureClient & provideDiagnosisKeys & Provide the keys of confirmed cases retrieved from  server\\
GoogleExposureClient & getExposureInformation & Provides a list of ExposureInformation objects, from which you can gauge the level of risk of the exposure\\
%\bottomrule
\end{tabular}
\end{table*}

\vspace{0.05in}
\noindent \textbf{Account Creation.}. The contact-isolation attack required the creation of accounts to simulate users. However, most applications limit the number of accounts each person can create by requiring unique information(e.g., medical ID, phone number) from users. Table \ref{tab:required_info} shows the required information for applications that we have tested. Unfortunately, due to possible privacy issues, not all applications require user to provide sensitive unique information (e.g. passport number, medical ID). Instead, they use a phone number or device ID to identify users, which can be easily 
forged by attackers. In our evaluation, we successfully registered users in all applications in Table \ref{tab:required_info} except \emph{tracetogether}, which requires a valid ID in Singapore. \emph{Stop corona}, \emph{swissCovid}, and \emph{erouska} use the Google and Apple framework to gather anonymous ID, from other users, and device ID is the only required information to create accounts. Thus, we created multiple users by emulating Android phones using Bluestack. \emph{Mytrace}, \emph{aarogya}, \emph{rakning} and \emph{Covidsafe} require phone number to create account but the phone number will only be verified during the registration. Hence, we leveraged online temporary phone-number services to receive the tokens for registration and created accounts in the four applications.

\vspace{0.05in}
\noindent \textbf{Contact-Isolation Attack.} To simulate the contact-isolation attack, four devices (A, B, C, and D) are placed so that the attacker device (D) is the one and only device that communicate with all three other devices.  A, B, and C are all normal devices and the user of C is assumed to be Covid positive. The attacker's device D has close contact with the three devices. The goal of the attacker is to pinpoint  the confirmed case. As mentioned in Section \ref{sec:implementation}, D will create three accounts to communicate with A, B, and C (i.e., denoted as $V_A$, $V_B$, and $V_C$, respectively). Then D exchanges data with the three devices using the corresponding identity.
%as shown in Figure \ref{fig:attsimulation:b}. 
When all data has been exchanged, C reports its status to the server and other devices pull the report from the server. The data stored in each device/account are as shown in Table \ref{tab:recordeddata}. The recorded encounters of $V_A$, $V_B$, and $V_C$ are assigned by our tool, according to Algorithm \ref{alg:contactanalyzer}. The results show that only data from C can be matched to the database from the server. Hence, C is the confirmed case, and D can obtain the corresponding device information from its log file.

\vspace{0.05in}
\noindent \textbf{Contact-Pollution Attack.} We leveraged the Bluetooth advertisement message to simulate the contact-pollution attack. The Bluetooth advertisement message is the most popular communication tunnel for CTAs. In our evaluation, we implemented a tool for Mac OS to automatically broadcast the advertisement messages it receives. %The devices are placed as shown in Figure \ref{fig:attsimulation:a}.
The four devices are carefully placed such that A, B, and C can all communicate with D but are out of range to communicate with each other. Since D will replay the messages it receives, $c_{A}, c_{B}$ and $c_{C},$ will be broadcast. Hence, A, B, and C will receive the messages and record the message since they are within the specific range of D. Then we use C to report its positive diagnostics to the server. The results show that both A and B received the notification messages although they did not have close contact with C.

The number of advertisement messages that a device can send at the same time is limited by its physical limitation. According to the Bluetooth specification, the maximum time interval between two messages is four seconds. For most applications, the duration for recording a close contact is at least five minutes, which means we need to continuously send the advertisement message during the five minute period. Each time we send an advertisement message, we need about 100 milliseconds (varies across devices). Summing  the time cost of other logic (including regenerating and capturing messages) in the application, adding an advertisement message into our list will cost around 400 milliseconds. Thus, we can target at most $4000 / 400 = 10$ devices at the same time with one attack device. Although the number of victims are limited, the distance between attackers and victims can be much larger. By leveraging commodity long-range Bluetooth transmitters, attackers may be able to launch attacks from as far away as 100 meters.

\begin{figure*}[htbp]
\centering
    \subfigure[Encounter Count]{
        \label{fig:visit_encounter}
        \begin{minipage}[t]{0.6\columnwidth}
        \centering
        \includegraphics[width=1.1\columnwidth]{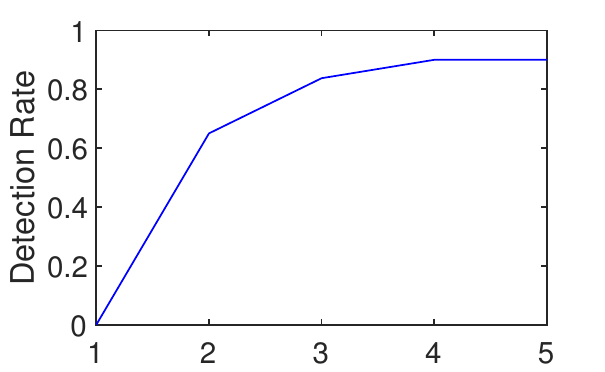}
        \end{minipage}
    }
    \subfigure[WiFi Probing Frequency]{
        \label{fig:wifi_fre}
        \begin{minipage}[t]{0.6\columnwidth}
        \centering
        \includegraphics[width=1.1\columnwidth]{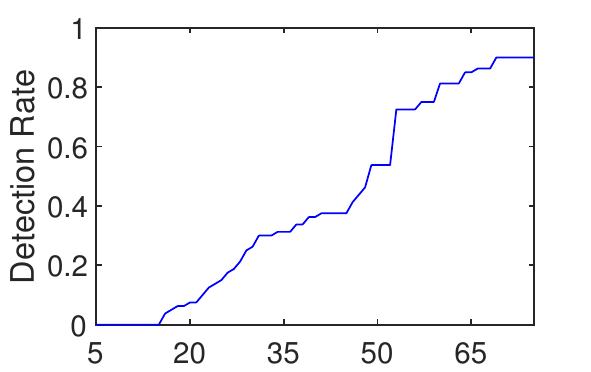}
        \end{minipage}
    }
    \subfigure[Speed]{
        \label{fig:speed}
        \begin{minipage}[t]{0.6\columnwidth}
        \centering
        \includegraphics[width=1.1\columnwidth]{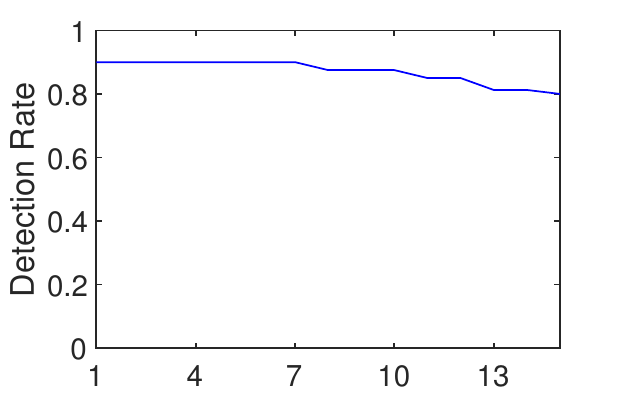}
        %\caption{fig2}
        \end{minipage}
    }%
 \vspace{-0.1in}
 \caption{Change in detection rates when the specific parameter for infected tags [(a): encounter count, (b):wifi\_frequency (c): speed] vary according to values shown in the X-axis.  In each case, all other parameters for both infected tags and normal tags are chosen from a uniform distribution, as described in Table~\ref{tab:params}.}
\label{fig:detection_rate}
\end{figure*}

\subsection{Simulating Large-Scale Deanonymization}
\label{sec:de-ano}
Since devices often randomize their Bluetooth MAC addresses, it is still hard to identify the real person with the Bluetooth MAC address obtained by the basic contact-isolation attack. Hence, we evaluate how we can further de-anonymize user devices by associating WiFi probe messages with observed Bluetooth MAC addresses ($M_{bt}$).  Our objective here is to move from an ephemeral identifier to a more persistent identifier, (i.e., WiFi MAC address and SSID tuple).

\begin{table*}[!h]
    \centering
    \caption{Summary of simulation parameters that were used in the large-scale deanonymization experiment.}
    \label{tab:params}
    \small
    \begin{tabular}{cll}
    %\toprule
    Parameter & Description & Value \\
    \midrule
    
    Population & Total user population in the simulation & 10000\\
    Infection Rate & Percent of infected users in the simulation & 1\%\\
    Time & Duration of the experiment & 16:00-18:00, 2 days\\
    Duration & Total time that each user stays in the range of sensors & 30-300 seconds\\
    Encounter Count & Expected encounter count of users & 1-5\\
    \# of tags & Total number of unique tags & 12000\\
    \# of infected tags & Number of unique tags that belong to infected users & 80\\
    % a more precise name
    Frequency(WiFi Probing) & WiFi probing rate of user devices & 15-75 probes per hour\\
    Frequency(Bluetooth) & Bluetooth advertisement message rate of user devices & 3-10 messages per second\\
    Speed & Mobility of users & 1-15 m/s \\
    %Movement & The movement strategy of users & Straight, Circle\\
    $r_{wifi}$ & Detection radius of WiFi Probe messages & 50 m\\
    $r_{bt}$ & Detection radius of Bluetooth advertisement messages & 10 m\\
    %$s$ & The average speed of users & 3 m/s\\
    %\bottomrule
    \end{tabular}
\end{table*}

A WiFi probe message is sent when WiFi is enabled but no connection has yet been established. These messages are typically broadcast with the name of recently connected SSIDs. Since most private SSIDs are unique, we can use a tag, $(M_{wifi}, SSID)$, to uniquely identify user devices if 
we can associate $M_{bt}$ with the corresponding tag. 
Both iOS and Android will randomize $M_{wifi}$ to preserve user privacy. However, the implementation randomizes the MAC address only for different SSIDs (i.e., the MAC address will be consistent with respect to specific SSIDs for extended periods). To associate $M_{bt}$ with the tag of the same device, we can monitor the entry and exit time of devices to find the possible pairs and further reconfirm the exact association when they subsequently reappear.

Based on the assumptions above, we designed the following approach to automatically identify the tag, $tag_{inf}$, of a given $M_{bt}$ that has been identified to be an infected user.

\begin{itemize}
    \item[1.] {\bf Tag Collection.} For each $M_{bt}$ detected at $t_1$, we consider all tags detected $2*(r_{wifi}-r_{bt})/s$ seconds before $t_1$. Here, $r_{wifi}$ and $r_{bt}$ represent the detection radius of WiFi probe messages and Bluetooth advertisement messages. $s$ is the average speed of users, which varies in different scenarios. In other words, we consider all tags detected at $t_2$ if $t_2 \ge t_1 - 2*(r_{wifi}-r_{bt})/s$. All qualified tags will be stored in $T$ and we use $(M_{bt}, T)$ to record the encounter.
    \item[2.] {\bf Tag Matching.} When $M_{bt\_inf}$ of infected user is identified by the Contact-Isolation Attack, we perform retrospective tag matching for $(M_{bt\_inf}, T_{inf})$. The candidate tags corresponding to an infected user are determined as follows: 1) $tag_{cand} \in T_{inf}$, 2) $\forall (M_{bt},T), if\; T \supset \{tag_{cand}\}, M_{bt}$ must have been identified as belonging to an infected user. Finally, among all candidate tags that satisfy the requirements, $tag_{inf}$ is the one that appears most frequently.
\end{itemize}

The aforementioned approach works well if all the infected users appear multiple times and their WiFi modules are always enabled. If an infected user only appears once or his WiFi module is disabled, false positives could be introduced when another tag happens to be in $T_{inf}$, but never appears again. False negative cases may result from multiple reasons. For example, (1) users may have explicitly turned off WiFi probing or (2) stay in range for a very short amount of time.  Finally, if an infected user always happens to appear along with another unrelated normal user, we will not be able to narrow down to a single infected user. However, such cases are a small fraction of the dataset, and even in these cases our approach can still help us narrow down to a smaller suspicious user set.

We evaluated the feasibility of the approach in a simulated environment. Table \ref{tab:params} shows the parameters considered in the simulation whose range of values were derived from published literature. The frequency and radius values are defined based on the observations of existing works~\cite{freudiger2015talkative}~\cite{googleapple2020exposurenotif}~\cite{wiki_wifi}. Here, 10000 users (of which 100 are infected) move from the outside toward our attack devices at a constant speed, stay for several seconds, and then leave at the same speed. When the data is distributed evenly, we generated 26,874 $(M_{bt}, T)$ pairs in total and successfully identified 72 (out of 80) tags that belong to infected users.  For the remaining 20 infected users, we observed neither the Bluetooth nor WiFi MACs because they never came in range.  We also evaluated the effect of each parameter on the detection rate. We fixed the parameters of infected users to specific value and the parameters of other users are still evenly distributed. The simulation results are shown in Figure \ref{fig:detection_rate}.
We find that WiFi probing frequency can significantly affect the detection rate. The reason is that once the frequency is too low, it is likely that the attack device will miss the messages, especially when users are moving at a high speed. Another important factor is visit encounter. The eight false negatives occur when the tags show up only once and other tags in the same pair cannot all be  excluded. In this case, we cannot narrow down the suspicious infected users to the exact tag, but we can generate a smaller list of possible tags. Since the simulation does not incorporate information beyond what might be observed in the real environments, we believe that it demonstrates the practical viability of the attack. With the unique tag, attackers would be able to physically locate the user~\cite{vanhoef2016mac} (e.g. by leveraging WiFi location data published by Wigle~\cite{wigle}) or track the user in the future~\cite{cunche2014know}.  In addition, video feeds could be combined with Tag database information to obtain a picture of candidate victims who might then be de-anonymized using face detection search engines like PimEyes~\cite{pimeyes}.

\subsection{Empirical Data Analysis}
\label{sec:empirical}

We leverage the POI database from SafeGraph to approximate the number of users possibly affected by our attacks. We assume that attackers are able to deploy as many attack devices as they want at many locations.  
The relevant fields we can get from the database include {\em location\_name, brands, street\_address, latitude, longitude, city, raw\_counts} (number of unique visitors for the period), and {\em visits\_by\_day}.
We cannot get the coverage of cities by adding up all the visitor counts because of the possible overlap between different places. Unfortunately, due to privacy considerations, the relevant data of the overlap (e.g., unique ID of visitors) is not available.

Two considerations apply to the overlap: ($i$) overlap of visitors across days for the same sensor location and ($ii$) overlap across locations.  Based on a 2017 study from Sense360, the average number of monthly customer visits at a single location for popular chains ranged between 1 and 2.2, with McDonalds being the most popular at 2.19 visits per month~\cite{sense360}.
Table~\ref{tab:visitors} provides the summary of monthly customer frequency values for the most popular fast-food chains.

\begin{algorithm} [t]
\caption{Coverage Estimation Algorithm}
\label{alg:coverage}
\small
\begin{algorithmic}[1]
    \REQUIRE Dataset of place pattern $S$, name of the target city $N$, population in the given city $n$,  overlap ratio $p$ and radius that the overlap applies to $r$.
    \ENSURE Approximate coverage of victims in city $c$
    \STATE $L = \emptyset$
    \FOR{$i \in S$}
        \IF{$i.city == N$}
            \STATE $L = L \bigcup {i}$
        \ENDIF
    \ENDFOR

    \FOR{$i \in L$}
        \FOR{$j \in L$}
            \IF{$i \neq j$}
                \STATE $dist = haversine(i.lat, i.long, j.lat, j.long)$
                \IF{$dist \le r$}
                    \STATE $i.visitor\_counts = i.visitor\_counts - p/2 * min(i.raw\_counts, j.raw\_counts)$
                    \IF{$i.visitor\_counts \le 0$}
                        \STATE remove $i$ from $L$
                    \ENDIF
                \ENDIF
            \ENDIF
        \ENDFOR
    \ENDFOR

    \STATE $unique\_counts = 0$
    \FOR{$i \in L$}
        \STATE $unique\_counts = unique\_counts + i.visitor\_counts$
    \ENDFOR
    
    \STATE $c = unique\_counts / n$
\end{algorithmic}
\end{algorithm}

On the other hand, by comparing the {\em raw\_counts} and the sum of {\em visits\_by\_day}, we can obtain the overlap across days in a given month for each location. Based on the dataset from May 2020 to June 2020, the average overlap for locations in the United States across days is 0.3674 ($1-raw\_visitor\_counts/\Sigma(visits\_by\_day)$). That means, that the average frequency of the SafeGraph visitor at each location ($\Sigma(visits\_by\_day)/raw\_visitor\_counts$) was 1.58 during the month.

\begin{table}[t]
    \small
    \centering
    \caption{Customer loyalty rates (monthly visit frequency) for popular fast-food chains.}
    \begin{tabular}{cc} %\toprule
         {\bf Fast-Food Chain} & {\bf Customer Frequency (Monthly)}  \\
         \midrule
         McDonalds & 2.19 \\
         Starbucks & 1.97 \\
         Sonic & 1.83 \\
         Tim Horton's & 1.82 \\
         Dunkin Donuts & 1.79 \\
         Whataburger & 1.74 \\ 
         %\bottomrule
    \end{tabular}
    \label{tab:visitors}
    \vspace{-0.2in}
\end{table}

\begin{figure*}[htbp]
\centering
    \subfigure[Houston]{
        \label{fig:houston}
        \begin{minipage}[t]{0.63\columnwidth}
        \centering
        \includegraphics[width=1.1\columnwidth]{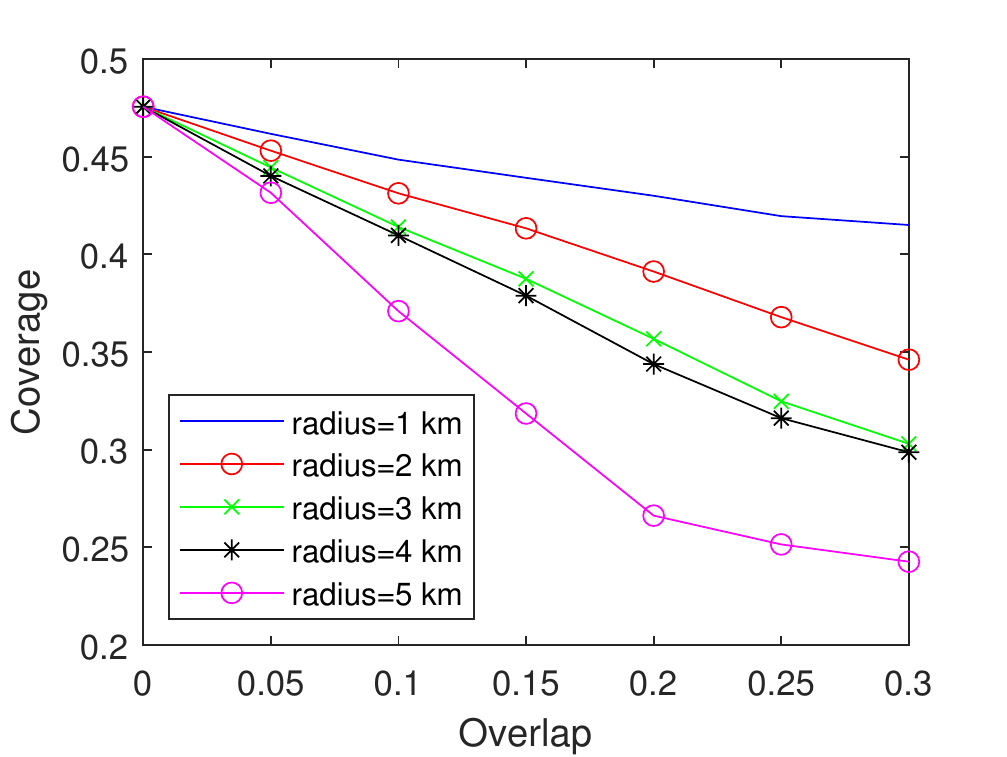}
        \end{minipage}
    }
    \subfigure[Bryan]{
        \label{fig:bryan}
        \begin{minipage}[t]{0.63\columnwidth}
        \centering
        \includegraphics[width=1.1\columnwidth]{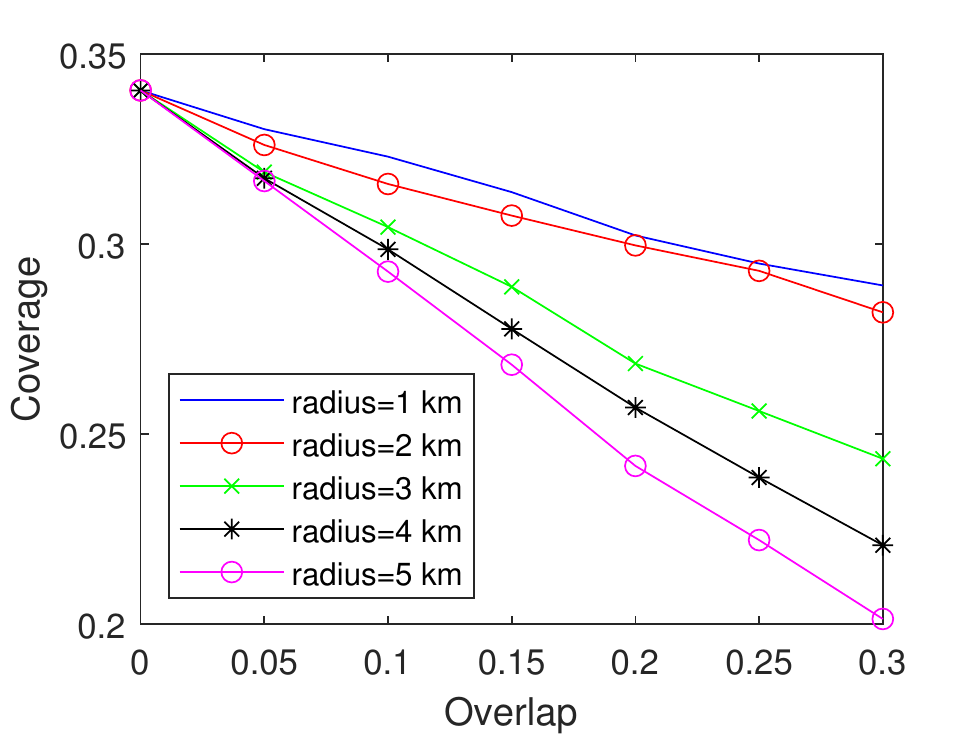}
        \end{minipage}
    }
    \subfigure[Chicago]{
        \label{fig:chicago}
        \begin{minipage}[t]{0.63\columnwidth}
        \centering
        \includegraphics[width=1.1\columnwidth]{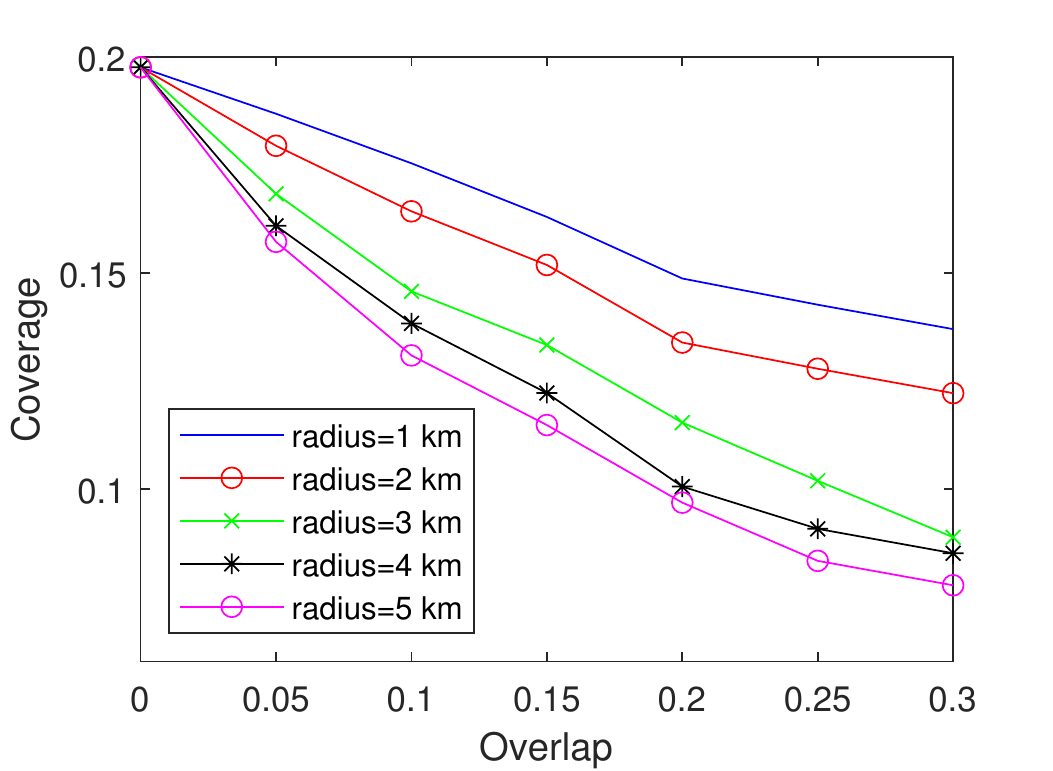}
        %\caption{fig2}
        \end{minipage}
    }%
    \vspace{-0.1in}
\caption{Fraction of potential victims tracked using SafeGraph sensors in different cities, assuming varying rates of overlap between sensors within a certain radius.  Houston (10,171 sensors), Bryan (446 sensors), Chicago (7,675 sensors).}
\vspace{-0.15in}
\end{figure*}

To address the cross-location overlap problem, we assume different ratio of overlap with different radius of the places and try to estimate the coverage of cities. For example, \emph{\{overlap:0.05, radius: 5\}} means that we consider each two places of which the distance is 5 miles and the overlap between them is 5\%. We use Haversine formula~\cite{robusto1957cosine} to calculate the distance between each two places from the coordinates. The algorithm of calculating the coverage is as shown in Algorithm \ref{alg:coverage}.
In our evaluation, we tested the coverage for three cities: Houston, TX; Bryan, TX; and Chicago, IL. To calculate the coverage, we got the population of the two cities from the United States Census Bureau\cite{censusbureau}. For Houston, the estimate population is 2,326,090. The coverage of possible victims with different overlap ratio and radii is illustrated in Figures~\ref{fig:houston}, \ref{fig:bryan}, and \ref{fig:chicago}. The results show that the attacks can possibly affect a relatively large portion of people. Even when we assume the visitor overlap for each two sensors within a 5 km range is as high as 30\%, the coverage of possible victims is still about 25\%.  For Chicago, this number drops to 5\%, which is still significant.   From the analysis of SafeGraph data from these three cities, we can see that our attack is effective in both big cities and small towns.

While it is difficult to get exact numbers for overlap and coverage, a slight variance in these results does not fundamentally discount the potential of the attack.  More importantly, they have implications for how the system is engineered.  For example, a lower overlap number implies that coverage of the attack is higher; and implies that we may have fewer opportunities to narrow down the victim set using pool testing. Since the contact-isolation attack relies on encounters and not visits, and each visit might have multiple encounters, the de-anonymization attack may be successful with a single visit.  On the other hand, a higher overlap number means less coverage but more opportunities to winnow down using pool-testing strategies. Specifically, a lower overlap corresponds to more victims of contact-pollution attack and a higher overlap corresponds to higher success rate in contact-isolation attack.

\section{Related Work}
\label{sec:related}

This paper is not the first to uncover security flaws in CTAs.  However, to our knowledge this is the first academic work to systematically analyze CTAs and point out the use of device simulation, pool-testing,and online databases to enable large-scale de-anonymization of CTA users.  We also raise awareness of the vulnerability of contact-tracing databases to data pollution attacks as well as the potential of Bluetooth range extenders in scaling such attacks.  To our knowledge, we are also the first to use the SafeGraph database to assess the significance of contact-tracing attacks.  The following summarizes and acknowledges related prior work that informed our research across various domains, including social media, Bluetooth, and contact-tracing protocols. 

\vspace{0.05in}

\noindent \textbf{SafeGraph for Human Cluster Tracking.}  Our paper explores the use of the SafeGraph POI dataset ~\cite{SafegraphDataset2020} as a resource for identifying the placement of contact-isolation attack devices to maximize engagement with contact-tracing mobile apps. This use of SafeGraph is quite consistent with other academic studies that examine the impact that Covid-19 has had on patterns of human movement. A number of live Covid-19 tracking projects (dashboards and interpretive studies) have emerged, including studies of fine-grained shelter-in-place compliance ~\cite{FootTraffic2020,bushman2020}, 
social-distancing compliance by region and political view~\cite{NBERw27531,Cook2020,SafegraphPolitics2020}, and the impact of this pandemic on categories of retail business~\cite{FootTraffic2020}.  Government policy studies on the effects of uncoordinated Covid-19 social-distance measures have also employed SafeGraph~\cite{Holtz2020}.  These studies use the mobile device geographic reporting information anonymized within the SafeGraph POI dataset to measure human foot-traffic patterns. This paper uses this information to identify human cluster points in a region to maximize the number of potential CTAs that a stationary attack device would reach.

\vspace{0.05in}

\noindent \textbf{Bluetooth Security.}
As one of the most popular wireless communication protocols, the security of Bluetooth has been a hot topic for security researchers. Ben et al.\cite{seri2017blueborne} found eight vulnerabilities in the implementation of Bluetooth on different platforms, which are named Blueborne. Daniele et al.~\cite{antonioli2020bias} proposed Bluetooth impersonation Attacks (BIAS) which allow attackers to impersonate other devices. BIAS abuses the switch-role message in the Bluetooth protocol to bypass the authentication process. The BlueFrag exploit (CVE-2020-0022) allows a remote attacker within proximity to silently execute arbitrary code with the privileges of the Bluetooth daemon~\cite{bluefrag}.

\vspace{0.05in}

\noindent \textbf{Contact Tracing.} 
To establish a common privacy-preserving standard for 
contact-tracing protocols, Google and Apple announced a two-phase exposure notification solution on April 10, 2020~\cite{googleapple2020exposurenotif}. Their solution provides a set of API for developing CTAs. The specifications and code are all available from the official website. Almost at the same time, an independent group started by academics in EPFL released DP-3T~\cite{troncoso2020decentralized}, which is fully open-sourced. We built and evaluated our prototype attacks on the DP-3T platform, but our attacks are generic and can be extended to these platforms.

Lars et al. show how the Google/Apple contact-tracing framework and DP-3T could be abused to physically locate infected persons in a large metropolitan city~\cite{baumgartner2020mind}. However, their de-anonymization technique differs from ours in that it relies on continuous tracking of Bluetooth activity from infected users across the city from multiple vantage points and does not leverage device spoofing, WiFi information, or online databases. In contrast, our attack only requires a single vantage point. Thus, the two efforts are complementary and could be combined for greater efficacy.  Serge~\cite{vaudenay2020analysis} also find that DP-3T is vulnerable to relay and replay attacks, which can introduce false alerts. However, this work does not actually demonstrate the feasibility of the attack and their proposed mitigation requires interactions between devices, which is not supported for Bluetooth advertisement messages. Krzysztof~\cite{pietrzak2020delayed} proposed a more practical solution by explicitly also checking for time and location. Similarly, the PACT protocol~\cite{rivest2020pact} uses a seed and timestamp to generate encounter chirps. 
Both solutions improve resilience against the contact-pollution attack but are still vulnerable to contact-isolation attack.  In addition, these mitigation may not always work against contact-pollution attacks, as sharing location with other devices is not always available and checking encounter time is insufficient to completely defend against contact-pollution attacks (e.g., the attacker may still broadcast recent encounter codes collected from other users).  Finally, there have been recent academic efforts such as Epione~\cite{epione}, which uses cryptographic techniques such as private set intersection cardinality (PSI-CA) and private information retrieval (PIR) to protect user privacy. Our proposed contact-isolation and contact-pollution attacks are fundamental (much like DDoS) and also affect this system.  Section~\ref{sec:countermeasures} discusses some potential countermeasures that could be deployed to deal with such attacks.
\section{Discussion}
\label{sec:discussion}
\label{sec:countermeasures}

This section discusses the ethical considerations of this work,  potential countermeasures for the two attacks presented in the paper,
and some limitations of the adversary models.

\subsection{Ethical Considerations}
A full evaluation of attack efficacy and our analysis of deployment challenges using existing live deployed contact-tracing systems raises prohibitive ethical concerns.  Such evaluations could hinder server operations, pollute contact-tracing databases, and cause false notification messages to be sent to users.  Therefore, we do not propose the execution of experiments that interact with fielded contract-tracing services.  Rather, the evaluation approach presented in Section~\ref{sec:eval} involves experimental devices that are divided into an attack set and a victim set, in which neither device set has contact with non-evaluation devices.  All adversarial models evaluated in this paper were operated in isolation.

Another important ethical consideration is the appropriate strategy for vulnerability disclosure.  Given that the specific attacks affect a broad class of applications, there is not an efficient and optimal strategy for privately disseminating this information.  Instead, we intend to share the specific attacks with popular framework designers for further guidance.

\subsection{Attack Countermeasures}
\textbf{Identity Verification.} For the attacks presented in this paper, an adversary  will need to create multiple accounts for uploading data and receiving notification messages. 
Based on our observations, most applications do not impose strict validations that could counter  
an attacker from producing a large account sets.   
For example, Smittestopp only requires a phone number to create a new account, which can be obtained using online phone-number services. Hence, employing multiple factors to reduce or restrict acceptable verification credentials can reduce the attack population that drives these attacks. Unfortunately, due to the possible privacy issues, to implement such an approach is difficult, may reduce the desired adoption rate of the CTA, or may be illegal in some countries.

\vspace{0.05in}
\noindent \textbf{Integrity/Consistency Check.} An assumption of the attacks is that the server will not check the integrity of the data or the consistency between the data uploaded by users and its origin device. Though attackers are able to send fake data to other users or to the server, additional integrity and consistency checks could be applied to guarantee that the data-exchange process cannot be sniffed or spoofed.

\vspace{0.05in}
\noindent \textbf{Notification Restrictions.} To launch the contact-isolation attack, attackers need to receive notification for each subset $C_i$, which means they need to continuously request the notification with a small encounter list. To mitigate this attack, we can set restrictions on the server side before we send notifications to users. For example, we may require a minimum-length encounter list to receive notifications.
However, if this becomes widely deployed, attackers may attempt to evade this restriction by introducing fake (synthetic) encounters into their submissions. Therefore, incorporating a threshold-length encounter list should be coupled with integrity-consistency checks that vet and reject fake encounters.

\subsection{Attack Limitations}
We now discuss some limitations associated with our adversary models. First, our attacks require the deployment of devices that must achieve physical proximity to victims, which limits the scalability of the attacks to the number of devices that can be created, deployed, and maintained. The cost may be unacceptable if the attack goal is to introduce a large volume of fake data into the database. 
Second, our attacks are highly dependent on the application implementation.  While the adversary models presented here are broadly applicable to a breadth of CTA schemes, the attack implementations need to be individually tailored per CTA. Finally, our attacks are most effective in scenarios where a large portion of the user population runs a monolithic CTA.

\section{Conclusion}
\label{sec:conclusion}
The paper examines two adversary models that can be applied to a wide range of Covid-19 CTAs that are emerging across the globe and are even being legally mandated among numerous countries.  CTAs are broadly categorized into centralized coordination strategies, which raise inherent public privacy concerns, and decentralized peer-oriented strategies that claim to reduce end-user privacy concerns.  However, we show that our adversary models apply equally well to both the centralized and decentralized strategies.  The first adversary model, called contact-isolation attack, presents a privacy attack against app users who become infected with Covid-19, enabling an attacker to deploy probes that could essentially mine infected user identities at locations where humans commonly congregate.  This attack combines large-scale device spoofing with pool testing, device re-identification strategies, and online databases to de-anonymize infected user devices.  We evaluate the efficacy of this attack both using a simulation study and the SafeGraph POI dataset.  The second adversary model employs a data-poisoning attack that enables an adversary to undermine the reliability of the CTA, resulting in false positive alarms that would be highly disruptive to any contract-tracing user base. We also discuss some mitigation strategies that could reduce the efficacy of both attacks.


% \section*{Acknowledgment}




%\bibliographystyle{IEEEtranS}


% \end{thebibliography}

\bibliographystyle{plain}
\bibliography{reference}

\end{document}